\documentclass[a4paper,11pt]{article}

\usepackage{contribution}
\usepackage{graphicx}

\newcommand{\rts}{\mbox{$\sqrt{\mathrm{s}_{_{\mathrm{NN}}}}$}}
\newcommand{\DtoKpi}{${\rm D^0 \to K^-} \pi^+$}
\newcommand{\DtoKpipi}{${\rm D^+\to K^-}\pi^+\pi^+$}
\newcommand{\DstartoDpi}{${\rm D^{*+}(2010)\to D^0}\pi^+$}
\newcommand{\Dzero}{$\rm D^0$}
\newcommand{\Dstar}{$\rm D^{*+}$}
\newcommand{\Dplus}{$\rm D^+$}

\newcommand{\weblink}[2][]{%
    \ifthenelse{\equal{#1}{}}%
    {\textnormal{\url{#2}}}%
    {\textnormal{\href{#2}{#1}}}%
}

\def\beq{\begin{equation}}
\def\eeq#1{\label{#1}\end{equation}}
\def\eeqn{\end{equation}}

\def\beqa{\begin{eqnarray}}
\def\eeqa#1{\label{#1}\end{eqnarray}}
\def\eeqan{\end{eqnarray}}

\let\bar=\overbar

\def\Dslash{\not{\hbox{\kern-4pt $D$}}}
\def\dslash{\not{\hbox{\kern-2pt $\del$}}}

\def\msb{{\bar{\ssstyle M \kern -1pt S}}}

\newcommand{\contribution}[7][]{%
  \clearpage
  \thispagestyle{plain}
  \ifthenelse{\equal{#1}{}}
  {\hypersetup{pdftitle={#2}}}
  {\hypersetup{pdftitle={#1}}}
  \hypersetup{pdfauthor={{#3} {#4}}}
  {\centering\normalfont\LARGE\bfseries\sffamily #2 \par\nobreak}
  \lhead{}
  \chead{%
    \textit{\footnotesize XIV International Conference on Hadron Spectroscopy
      (\weblink[\textit{hadron2011}]{http://www.hadron2011.de}), 13-17 June 2011, Munich, Germany}%
  }
  \rhead{}
  \bigskip
  \begin{center}
    {#3} {#4}\ifthenelse{\equal{#6}{}}{}{\footnote{\weblink[#6]{mailto:#6}}}
    \ifthenelse{\equal{#7}{}}{}{#7} \\
    \textit{#5}
  \end{center}
  \bigskip
}

\renewcommand{\abstract}[1]{%
  \begin{center}
    \begin{minipage}{0.85\textwidth}
      \begin{footnotesize}
        #1
      \end{footnotesize}
    \end{minipage}
  \end{center}
  \bigskip
}

\begin{document}

%
%
%
%
%
{  


%

\contribution[]  
{Heavy-flavor production in pp and Pb--Pb collisions at LHC with ALICE }  
{Kai}{Schweda}  
{Physikalisches Institut der Universit\"at Heidelberg, Philosophenweg 12,\\
 D-69120 Heidelberg, Germany}  
{kschweda@cern.ch}  
{on behalf of the ALICE Collaboration}  
%

\abstract{
We report recent results from the ALICE experiment the Large Hadron Collider on open charmed hadron production in pp collisions 
at $\sqrt{s} = 2.76$ and 7~TeV, and Pb--Pb collisions at $\sqrt{\mathrm{s}_{_{\mathrm{NN}}}} = 2.76$ TeV.  
Open charmed hadrons are kinematically fully reconstructed in the
hadronic decay channels
\DtoKpi, \DtoKpipi, and \DstartoDpi
and identified through their invariant mass. Combinatorial background is minimized by selecting a displaced vertex topology. Inclusive charm and beauty production is measured by detecting electrons (muons) from semi-leptonic decays of open charmed and beauty hadrons in the central (forward) region. Comparison to results from state-of-the-art QCD calculations is given.
First results on nuclear modifications factors in Pb--Pb collisions from hadronic and semi-leptonic decays are presented. }
%

\section{Introduction}
The ALICE experiment~\cite{ALICE-experiment} was designed to study strongly interacting matter at the highest temperatures and energy densities available in the laboratory,  in high-energy nucleus-nucleus collisions at the Large Hadron Collider. Heavy quarks (charm and bottom) are unique probes for studies of bulk phenomena, due to their large mass relative to the temperature of the medium,  and are abundantly produced at LHC energies. Here, heavy-quark production rates in pp~collisions provide the essential baseline for such studies in heavy ion collisions. In addition, quantitative understanding of heavy-quark production is crucial in the search for new physics phenomena at the LHC, where heavy-quark production often comprises an important and irreducible background~\cite{higgs}.

For hard processes in nucleus-nucleus $(AA)$ collisions, in the absence of any nuclear medium effects their production rates are expected to scale with the number $\langle N_{\rm coll} \rangle$ of binary nucleon-nucleon collisions when compared to pp rates. The nuclear modification factor to quantify this relationship is defined as:
\begin{equation}
R_{AA}(p_T) = \frac{1}{\langle N_{\rm coll} \rangle} \cdot \frac{{\rm d}N_{AA} / {\rm d}p_T}{{\rm d}N_{pp} / {\rm d}p_T} =  \frac{1}{\langle T_{AA} \rangle} \cdot \frac{{\rm d}N_{AA} / {\rm d}p_T}{{\rm d} \sigma _{pp} / {\rm d}p_T} 
\end{equation}
which equals unity in case of no nuclear effects.
Here, $\langle T_{AA} \rangle$ is the average nuclear overlap function calculated in a Glauber model of the nucleus-nucleus collision geometry.

The heavy-quark detection performance of ALICE is described in detail in \cite{ALICE-PPR}. The main components 
for these measurements in the central region of ALICE
are the Inner Tracking System (ITS) 
covering a radial distance of 3.9~cm to 43.0~cm  from the collision vertex and
based on high-granularity silicon
technology, surrounded by a Time Projection Chamber (TPC) embedded in a magnetic field of 0.5 T. 
Both components provide high-precision tracking of charged particles in the pseudo-rapidity range $|\eta| < 0.9$, with a 
relative momentum resolution of better than 4$\%$ at $p_T < 20$~GeV/$c$ and a pointing resolution to the collision vertex of better than 75(20)~$\mu$m at $p_T > 1(20)$~GeV/$c$ in the plane transverse to the beam direction. Particle identification is provided by the specific energy deposit dE/dx in the TPC gas, and time-of--flight. 
Electrons at 
transverse momenta below 6~GeV/$c$ are identified by a combination of  dE/dx and time-of--flight, and above 2~GeV/$c$ by the Transition Radiation Detector and in the Electromagnetic Calorimeter.
Muons are identified utilizing a frontal absorber with a thickness of ten interaction lengths $\lambda_I$, and a muon filter of  thickness $7\lambda_I$  covering the pseudo-rapidity range $-4 < \eta < -2.5$. 
Data were recorded using a minimum bias trigger selection which was defined by  a signal present in either of two scintillator hodoscopes positioned at forward and backward direction, or in the ITS of the central barrel.

\section{Calibrating the probe in pp collisions at $\sqrt{s}$ = 7~TeV}

Open charmed hadrons within $|y| < 0.5$ are fully reconstructed in the channels
\DtoKpi, \DtoKpipi, and \DstartoDpi
\begin{figure}[htb]
  \begin{center}
    \includegraphics[width=0.28\textwidth]{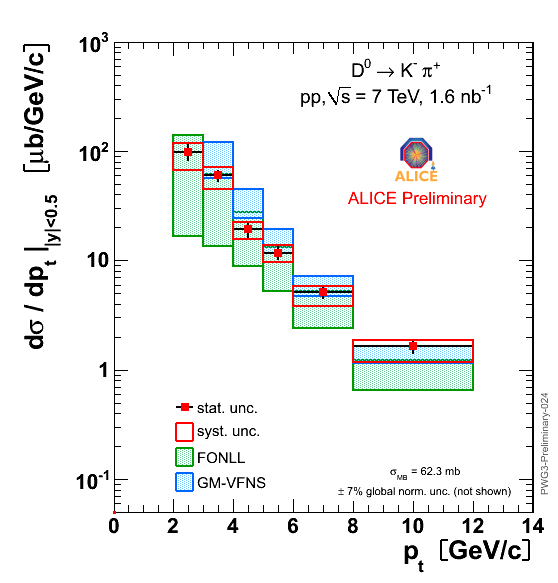}
    \includegraphics[width=0.28\textwidth]{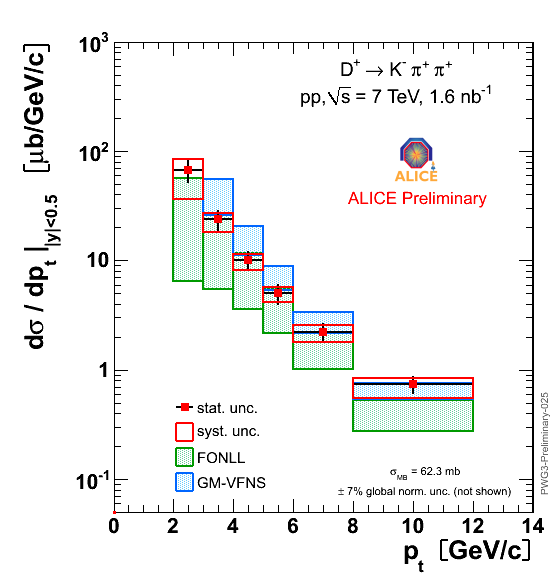}
    \includegraphics[width=0.28\textwidth]{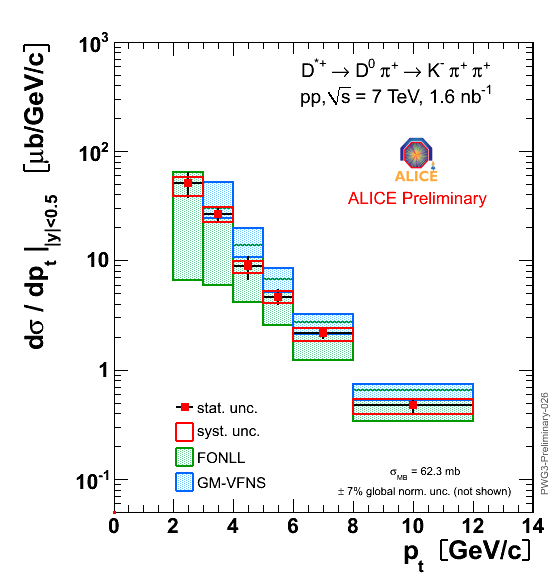}
    \caption{Differential production cross section of prompt \Dzero, \Dplus, and \Dstar mesons and their charge conjugates within the rapidity interval $|y| < 0.5$ in pp collisions at $\sqrt{s} = 7$~TeV and predictions from
    pQCD calculations~\cite{FONLL, GM-VFNS}.}
    \label{fig:D-pp}
  \end{center}
\end{figure}
 and their charge conjugates and identified through their invariant mass.
Combinatorial background is minimized by selecting a
displaced vertex topology, i.e. the separation of tracks stemming from the secondary vertex of the weakly decaying D meson that is displaced from the primary collision vertex. 
Particle identification, especially of
charged kaons, further reduces the background. The contribution of secondary D mesons from the decay of B mesons is corrected for, and amounts to about 15\% estimated
from  production cross sections using FONLL~\cite{FONLL} and applying reconstruction efficiencies obtained from detailed detector simulations.
The prompt D meson production cross section is shown in Fig.~\ref{fig:D-pp}.

The measured spectrum covers the transverse momentum range from 2 - 12 GeV/$c$ with an integrated luminosity of $\mathcal{L}_{\rm int} = 1.6\ {\rm nb}^{-1}$. With four times more statistics on tape from the year 2010 data taking, we expect to extend the momentum range down to 1 GeV/$c$ and up to at least 20 GeV/$c$. Results from calculations based on perturbative QCD~\cite{FONLL, GM-VFNS} are in agreement with measurement within uncertainties.
\begin{figure}[htb]
  \begin{center}
    \includegraphics[width=0.45\textwidth]{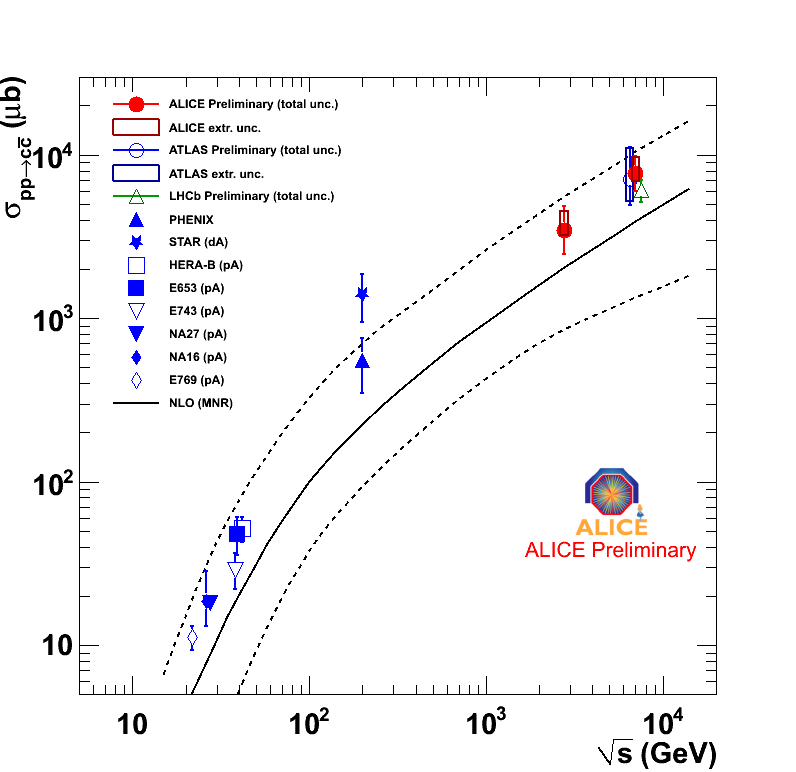}
    \caption{Open charm production cross section extrapolated to full phase space at various center-of-mass energies~\cite{Lourenco, ATLAS, LHCb, STAR, PHENIX}.}
    \label{fig:sigma-ccbar}
  \end{center}
\end{figure}

The measured cross sections were extrapolated to full phase space by scaling the measured cross section by  the ratio of the total cross section over the cross section in the experimentally covered phase space calculated with FONLL. Systematic uncertainties of the calculation were estimated as follows: the renormalization and factorization scale variables $\mu_R$ and $\mu_F$ were varied up and down by a factor of two with the additional constraint $0.5 < \mu_R / \mu_F < 2.0$, the charm quark mass was varied independently within $0.20$~GeV/$c^2$ around the central value at $m_c = 1.5$~GeV/$c^2$, uncertainties in the parton distribution functions were estimated using the CTEQ6.6 PDF error eigenvectors.  
The total charm production cross section was estimated for each species of D meson separately by dividing the total D meson production cross section by the relative production yield for a charm
quark hadronizing to a particular species of D meson.
The relative production yields have been measured at LEP 
at the Z-resonance~\cite{charm-frag,OPAL} and have been applied to our results.
We then calculated the weighted average of the total charm production cross section from the extrapolated values for  \Dzero , \Dplus , and  \Dstar.
The dependence of the total charm production cross section~\cite{Lourenco,ATLAS,LHCb,STAR,PHENIX} 
on the collision energy is shown in Fig.~\ref{fig:sigma-ccbar}. The error boxes around the ATLAS and ALICE points denote the extrapolation uncertainties alone, whilst the error bars are the overall uncertainties. 
The black curves show results from next-to-leading-order predictions from the MNR calculation framework, together with its uncertainties. 
Note that all data points populate the upper band of the theoretical prediction. 

Electrons from semi-leptonic decays of heavy quarks were extracted from the inclusive electron spectrum by subtracting a cocktail of background electrons with the dominant contributions from
$\pi^0$ and $\eta$ Dalitz decays and decays of the vector mesons $\rho, \omega$, and $\phi$ as well as photon conversions to electron-positron pairs in the detector material~\cite{Daines-qm11}.
The yields for neutral mesons were taken from the measured neutral pion cross section and the assumption of ${\rm m_T}$ scaling~\cite{mt-scaling}.  The inclusive raw electron spectrum has a residual
contamination from mis-identified charged pions of up to 15\% at high $p_T$, which has been determined experimentally and then subtracted . 
Muons were measured at forward rapidity in the muon spectrometer~\cite{Daines-qm11}. The inclusive muon spectrum contains three major background contributions: muons from the decay in-flight of light hadrons, 
muons from the decay of hadrons produced through interactions in the absorber, and hadrons punching through the front absorber. The latter background is efficiently rejected by requiring matching of tracks
in the spectrometer with tracks in the trigger system. The other two background sources are highly momentum dependent and were subtracted using results from detailed simulations. Our muon measurement 
covers the transverse momentum range from 2 - 10 GeV/$c$. Results from FONLL calculations are in good agreement with our data.

The collision energy for Pb--Pb was \rts = 2.76~TeV. To obtain the p+p reference cross section at this energy,
the measured pp cross section was scaled by the momentum dependent ratio of cross sections from FONLL calculations at $\sqrt{s}$ = 2.76 over 7~TeV~\cite{pp-scale}. The resulting cross section
reference for \Dzero\ and \Dplus\  was experimentally checked, though with limited precision, during a brief data taking run with pp collisions at $\sqrt{s}$ = 2.76~TeV and good agreement was found.

\section{Medium modifications in Pb--Pb collisions at \rts = 2.76 TeV}
In the analysis of 17 million minimum bias triggered Pb--Pb collisions, we followed an identical analysis strategy as outlined above for pp collisions. The collision centrality was defined 
by the sum of the signal amplitudes in the scintillator hodoscopes~\cite{Toia-qm11}.
The contribution of feed-down from B meson decays has been estimated from FONLL calculations
to be 10\%--15\%, depending on $p_T$, and was subtracted. 
The as-yet unknown nuclear modification of B mesons has been accounted for by varying the B meson yield up and down by a factor of 3, resulting in a
variation of the D meson nuclear modification factor of less than 15\%. Systematic uncertainties due to tracking, particle identification and topological selection of D mesons amount to about 35\%. The
nuclear modification factor of \Dzero\ (filled triangles) and \Dplus\ (filled squares) mesons is shown on the left side of  Fig.~\ref{fig:raa} for 0 - 20\% most central Pb--Pb collisions. We observe a strong suppression of a factor 4-5 for $p_T > 5$~GeV/$c$, comparable
with values for charged pions (filled circles).
\begin{figure}[htb]
  \begin{center}
    \includegraphics[width=0.33\textwidth]{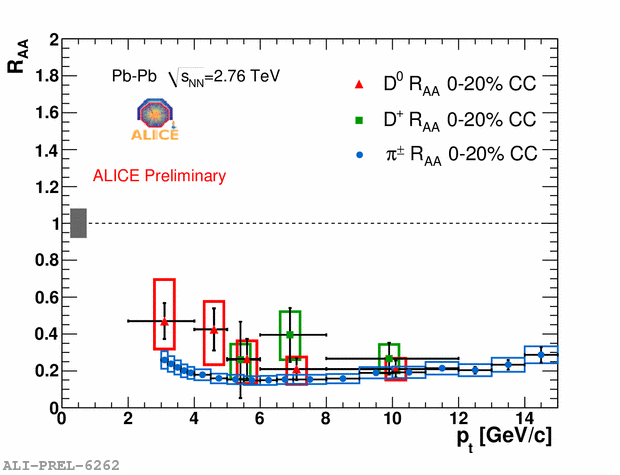}
    \includegraphics[width=0.29\textwidth]{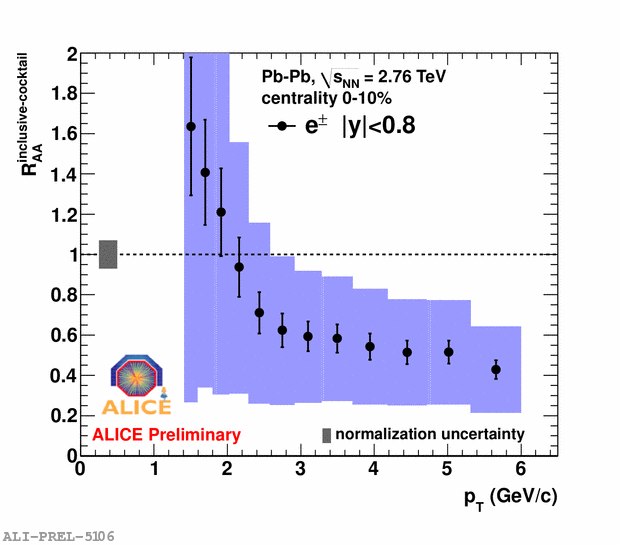}
      \includegraphics[width=0.33\textwidth]{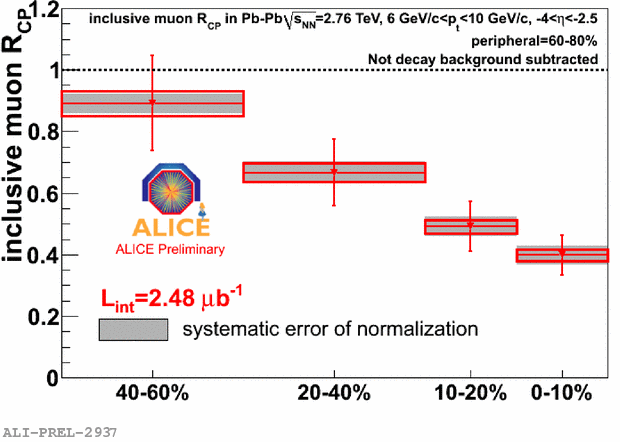}
     \caption{Left: Nuclear modification factor for $D^0, D^+$, and $\pi^+$ in central Pb--Pb collisions at \rts = 2.76~TeV. Statistical (bars), systematic (open boxes) normalization (filled boxes) 
     uncertainties are shown. Middle: Nuclear modification factor for electrons (filled circles) from semi-leptonic decays of heavy quarks. Right:  Nuclear modification factor for inclusive muons (filled triangles).}
    \label{fig:raa}
  \end{center}
\end{figure}
Electrons (filled circles) from semi-leptonic decays of heavy quarks show a suppression of a factor 1.5 - 4 above a transverse momentum of 4~GeV/$c$, as shown in the middle of Fig.~\ref{fig:raa}. 
The systematic uncertainty is dominated by the correction for the electron identification.
Inclusive muons (filled triangles), integrated over the  transverse momentum range from  $p_T > $ 6 to 10~GeV/$c$, are shown on the right side of Fig.~\ref{fig:raa} and exhibit a suppression
of a factor of larger than two in most central collisions. In this momentum range decays from B mesons should significantly contribute to the spectrum. We estimate the contamination from light-quark decays in the inclusive muon spectrum from detailed simulations to be less than 15\%.
The centrality dependence of the nuclear modification factor for D mesons, electrons from semi-leptonic decays of heavy quarks and inclusive muons shows a strong decrease towards more central collisions and is compatible with unity for most peripheral collisions.

\section{Conclusions}
ALICE has measured D mesons, and electrons and muons from semi-leptonic decays of heavy quarks in pp and Pb--Pb collisions at LHC. We observe a strong yield suppression in Pb--Pb collisions of a factor 4-5 for D mesons when compared to scaled results from pp collisions, comparable with the measured suppression for charged pions. Electrons from semi-leptonic decays of heavy quarks and inclusive muons also show
strong suppression of a factor up to 3 at momenta $p_T > 5$~GeV/$c$ where decays from B mesons are expected as the dominant contribution.
The upcoming high luminosity Pb--Pb run at the end of 2011 and a possible p--Pb run the year after will enable ALICE to precisely measure heavy-quark production, disentangle bottom from charm production and quantify initial-state nuclear effects.


%

}  


\end{document}